\documentclass[epj]{svjour}[24.12.01]
\usepackage{epsfig}
\usepackage{graphics}
\usepackage{amssymb}
\usepackage{dcolumn}
\usepackage{amsmath}
\usepackage{tabularx}
\usepackage{units}
\usepackage{changebar}

\begin{document}
\newcommand{\al}{\bar{\alpha}}
\newcommand{\be}{\bar{\beta}}
\newcommand*\prim{\,^\prime}
\newcommand*\thom{\frac{e^2}{m\,}\,}
\newcommand*{\textfrac}[2]{\ensuremath{\kern.1em
\raise.5ex\hbox{\the\scriptfont0 #1}\kern-.1em
/\kern-.15em\lower.25ex\hbox{\the\scriptfont0 #2}}}

\title{Quasi-free Compton Scattering and the Polarizabilities of the
  Neutron\thanks{Supported by Deutsche
    Forschungsgemeinschaft (SFB\,201, SFB\,443, Schwerpunktprogramm
    1034 through contracts DFG-Wi1198 and DFG-Schu222), and by the
    German Russian exchange program 436 RUS 113/510.)}}

\author{K. Kossert\inst{1}\thanks{Part of the Doctoral  Thesis}$^,$
\!\!\thanks{\textit{Present address}:
Physikalisch-Technische Bundesanstalt,
Bundesallee 100, D-38116 Braunschweig} 
\and
M. Camen\inst{1}\thanks{Part of the Doctoral Thesis} \and
F. Wissmann\inst{1}$^{\rm b}$\!$^,$\thanks{Part of the
Habilitation  Thesis} \and
J. Ahrens\inst{2} \and
J.R.M. Annand\inst{3} \and
H.-J. Arends\inst{2} \and
R. Beck\inst{2} \and
G. Caselotti\inst{2} \and
P. Grabmayr\inst{4} \and
O. Jahn\inst{2} \and
P. Jennewein\inst{2} \and
M.I. Levchuk\inst{5} \and
A.I. L'vov\inst{6}   \and
J.C. McGeorge\inst{3} \and
A. Natter\inst{4} \and
V. Olmos de Le\'on\inst{2} \and
V.A. Petrun'kin\inst{6} \and
G. Rosner\inst{3} \and
M. Schumacher\inst{1} \and
B. Seitz\inst{1} \and
F. Smend\inst{1}
A. Thomas\inst{2} \and
W. Weihofen\inst{1} \and
F. Zapadtka\inst{1}
}
\institute{II\@. Physikalisches Institut, Universit\"at G\"ottingen, D-37073
G\"ottingen, Germany \and
Institut f\"ur Kernphysik, Universit\"at Mainz,
D-55099 Mainz, Germany \and
Department of Physics and Astronomy, University of Glasgow,
Glasgow G12 8QQ, UK \and
Physikalisches Institut, Universit\"at T\"ubingen,
D-72076 T\"ubingen, Germany \and
B.I. Stepanov Institute of Physics, Belarussian Academy of
Sciences, 220072 Minsk, Belarus \and
P.N. Lebedev Physical Institute, 117924 Moscow, Russia 
} 
\date{Received: date / Revised version: date}

\abstract{Differential cross-sections for quasi-free Compton
  scattering from the proton and neutron bound in the deuteron have
  been measured 
  using the Glasgow/Mainz photon tagging spectrometer
  at the Mainz MAMI accelerator together with  the Mainz
  \unit[48]{cm} $\O$ $\times$ \unit[64]{cm} NaI(Tl) photon detector
  and the G\"ottingen SENECA recoil detector. The data cover photon
  energies ranging from \unit[200]{MeV} to \unit[400]{MeV} at
  $\theta^{\rm LAB}_\gamma = 136.2^\circ$. Liquid deuterium and
  hydrogen targets allowed  direct comparison of free and
  quasi-free scattering from the proton.  The neutron
  detection efficiency of the SENECA detector was measured via the
  reaction $p(\gamma,\pi^+ n)$.  The  "free" proton Compton scattering
  cross sections  extracted from the bound proton data are in
  reasonable agreement with those for the free proton which gives confidence
  in the method  to extract the differential cross section for
  free scattering from quasi-free data.  Differential cross-sections
  on the free neutron have been extracted and the difference of the
  electromagnetic polarizabilities of the neutron has been determined
  to be $\alpha_n - \beta_n = 9.8 \pm 3.6 \text{(stat)}
  \, {}^{+2.1}_{-1.1} \text{(syst)} \pm 2.2 \text{(model)}$ in units of
  $\unit[10^{-4}]{fm^3}$.  In combination with the polarizability sum
  $\alpha_n + \beta_n = 15.2 \pm 0.5$ deduced from photoabsorption
  data, the neutron electric and magnetic polarizabilities, $\alpha_n
  = 12.5 \pm 1.8 \text{(stat)} \, {}^{+1.1}_{-0.6} \text{(syst)} \pm 1.1
  \text{(model)}$ and $\beta_n = 2.7 \mp 1.8 \text{(stat)}
  \, {}^{+0.6}_{-1.1} \text{(syst)} \mp 1.1 \text{(model)}$ 
are obtained. The backward spin polarizability of the neutron
  was determined to be 
 $\gamma^{(n)}_\pi= (58.6 \pm 4.0) \times 10^{-4}{\rm fm}^4$.
 }


\maketitle

\section{Introduction}                                                        
The electromagnetic structure of the nucleon is a fascinating field of
current research. At low energies, the electromagnetic structure may
be parametrized by amplitudes for single-pion photoproduction
\cite{arndt96,drechsel99} and by the invariant amplitudes for Compton
scattering \cite{LPS97}.  Instead of using these amplitudes in their
general form, it has become customary to consider special properties of
these amplitudes which have a transparent physical interpretation.
These properties are given in terms of electromagnetic structure
constants of which the E2/M1 ratio of the $p \to \Delta$ transition,
the electric and magnetic polarizabilities $\alpha$ and $\beta$,
respectively, and the spin polarizabilities $\gamma_0$ and
$\gamma_\pi$, for the forward and backward direction, respectively,
are the most prominent.

These electromagnetic structure constants have been studied for the
proton  for a long time, whereas the corresponding
investigations for the neutron are only  beginning. This
contrasts with the fact that for their interpretation it is of great
interest to know whether or not the proton and the neutron have the
same or different electromagnetic structure constants.  This is one
reason to study the electromagnetic polarizabilities of the neutron in
addition to those for the proton. Since Compton scattering experiments
appeared too difficult, the first generation of investigations
concentrated on the method of electromagnetic scattering of low-energy
neutrons in the Coulomb  field of heavy nuclei, investigated in
narrow-beam neutron transmission experiments. The history of these
studies is summarized in Refs. \cite{aleksandrov92,aleksandrov01}. 
The latest in a
series of experiments have been carried out at Oak Ridge
\cite{schmiedmayer91} and Munich \cite{koester95} leading to
\begin{equation}
  \alpha_n = 12.6 \pm 1.5 \pm 2.0 
  \label{schmiedmayer}
\end{equation}
and
\begin{equation}
  \alpha_n = 0.6 \pm 5,
  \label{koester}
\end{equation}
respectively, in units of $\unit[10^{-4}]{fm^3}$ which will be used
throughout in the following. The numbers given here have been 
corrected by adding the Schwinger term \cite{lvov93} 
$e^2\kappa^2_n/4M^3 = 0.6$, containing the neutron anomalous magnetic moment 
$\kappa_n$ and the neutron mass $M$. This term had been omitted in the original
evaluation of these experiments. After including the Schwinger term 
the numbers are directly comparable with the ones defined through 
Compton scattering.
This means that the difference between the electric polarizabilities
obtained by the two methods, i.e. electromagnetic  scattering 
of neutrons in a Coulomb field
and Compton scattering, was only due to the incomplete formula used for the
evaluation of the electric polarizabilities derived from electromagnetic
scattering experiments, whereas the distinction between a Compton
polarizability $\bar \alpha$ and a "true" polarizability $\alpha_0$ was
not justified. The main argument is that $\alpha_0$ is not an
observable but merely a model quantity which requires great
precautions when used in model calculations \cite{lee01a,lee01b,lee02}. 
This clarification makes it unnecessary to use different notations
for the electric and magnetic polarizabilities depending on the definition. 
Therefore, we decided to use the simplest choice, i.e. $\alpha$ and $\beta$.

While the Munich result
\cite{koester95} has a large error, the Oak Ridge result \cite{schmiedmayer91}
is of very high precision.  However, this high precision has been
questioned by a number of researchers  active in the field of
neutron scattering \cite{enik97}. Their conclusion is that the Oak
Ridge result \cite{schmiedmayer91} possibly might be quoted as $7 \leq
\alpha_n \leq 19$. Furthermore, it should be noted that electromagnetic 
scattering of neutrons in a Coulomb field does  not constrain the 
magnetic polarizability $\beta_n$.

A pioneering experiment on Compton scattering by the neutron had been
carried out by the G\"ottingen and Mainz groups using photons 
produced by the  electron
beam of MAMI~A operated at \unit[130]{MeV} \cite{rose90}. This
experiment followed a proposal of Ref. \cite{levchuk94} to exploit
the reaction $\gamma d \to \gamma np$ in the quasi-free kinematics,
though there is an evident reason why such an experiment is difficult
at energies below pion threshold. For the proton the largest portion
of the polarizability-dependent cross section in this energy region
stems from the interference term between the Born amplitude containing
Thomson scattering as the largest contribution, and the non-Born
amplitude containing the polarizabilities. For the neutron the Thomson
amplitude vanishes so that the interference term is very small and
correspondingly cannot be used for the determination of the neutron
polarizabilities.  This implies that the cross section is rather small
being about \unit[2--3]{nb/sr} at \unit[100]{MeV}. The way chosen to
overcome this problem was to use a high flux of bremsstrahlung without
tagging \cite{rose90,levchuk94}.  The result obtained in the
experiment \cite{rose90} was
\begin{equation}
  \alpha_n=10.7 \, {}^{+3.3}_{-10.7}.
  \label{rose} 
\end{equation}
This means that the experiment was successful in providing a value for
the electric polarizability and its upper limit but it did not 
determine a definite lower limit. The reason for this deficiency is
that below pion threshold the neutron Compton cross section is
practically independent of $\alpha_n$ if $0 \lesssim \alpha_n \lesssim
10$ \cite{levchuk94,wissmann98}. In order to overcome this difficulty
it was proposed to measure the neutron polarizabilities at energies
above pion threshold with the energy range from \unit[200] to
\unit[300]{MeV} being the most promising since there the cross
sections are very sensitive to $\alpha_n$ \cite{levchuk94,wissmann98}.
In principle, this method allows a separate extraction of 
$\alpha_n$ and $\beta_n$. However, it is more precise to use the 
Baldin sum rule prediction for the sum of the nucleon polarizabilities.
A recent evaluation of this sum rule \cite{levc00} gives
\begin{eqnarray}
&\alpha_p + \beta_p = 14.0 \pm 0.3,&
\label{baldinp}\\
&\alpha_n + \beta_n = 15.2 \pm 0.5.&
\label{baldinn}
\end{eqnarray}

A first experiment on quasi-free Compton scattering by the proton
bound in the deuteron for energies above pion threshold was carried
out at MAMI (Mainz) \cite{wissmann99}. This experiment served as a
successful test of the method of quasi-free Compton scattering for
determining $\alpha_n - \beta_n$. The result obtained form the quasi-free
data, $\alpha_p - \beta_p = 9.1 \pm 1.7 \text{(stat+syst)} \pm 1.2
\text{(model)}$ proved to be in reasonable agreement with the world
average of the free-proton data which is $\alpha_p - \beta_p = 10.5
\pm 0.9 \text{(stat+syst)} \pm 0.7 \text{(model)}$ according to the
most recent analysis \cite{olmos01}. Later on this method was applied
to the proton and the neutron bound in the deuteron at SAL (Saskatoon)
\cite{kolb00}. In this experiment differential cross sections for
quasi-free Compton scattering by the proton and by the neutron were
obtained at a scattering angle of $\theta^\text{LAB}_\gamma =
135^\circ$ for incident photon energies of \unit[236 -- 260]{MeV}, which
were combined to give one data point of reasonable precision for each
nucleon. From the ratio of these two differential cross sections a
most probable value of
\begin{equation}
  \alpha_n - \beta_n = 12
  \label{kolb1}
\end{equation}
was obtained with a lower limit of 0 and no definite upper limit.
Combining their result \cite{kolb00} with that of Eq. (\ref{rose})
\cite{rose90} the authors obtained the following 1$\sigma$ constraints
for the electromagnetic polarizabilities
\begin{equation}
  7.6 \leq \alpha_n \leq 14.0
\label{kolb2}
\end{equation}
and
\begin{equation}
  1.2 \leq \beta_n \leq 7.6.
\label{kolb3}
\end{equation}

It should be noted that coherent elastic (Compton) scattering by
the deuteron provides a further method for determining the electromagnetic
polarizabilities of the neutron. Measurements of  differential
cross sections  of this process have been performed at Illinois \cite{lucas94},
SAL \cite{hornidge00}, and MAX-lab \cite{lundin02}. An evaluation of these 
experiments in the framework of the theoretical approach \cite{levc00} 
gives  the following values for the neutron polarizabilities 
\begin{eqnarray}
&\alpha_n=9.4\pm 2.5,& \quad \beta_n= 5.8\mp 2.5\quad {\rm (Illinois)},
\label{Illinois}\\
&\alpha_n=5.5\pm 2.0,& \quad \beta_n= 9.7\mp 2.0\quad {\rm (SAL)},
\label{SAL}\\
&\alpha_n=9.2\pm 2.2,& \quad \beta_n= 6.0\mp 2.2\quad {\rm (MAX-lab)}.
\label{MAX-lab}
\end{eqnarray}
These values are normalized to the Baldin sum rule prediction for the 
neutron (\ref{baldinn}). The electromagnetic polarizabilities for the
proton entering into the data evaluation are taken from
the respective Baldin sum rule (\ref{baldinp}) together with 
the currently accepted global average
$\alpha_p - \beta_p = 10.5$ \cite{olmos01}. The errors in Eqs.
(\ref{Illinois}) -- (\ref{MAX-lab}) are experimental only and do not contain
uncertainties related to the calculation of the nuclear-structure 
part of
the scattering amplitudes. Though such a model error could be estimated
for the method used in the calculation leading to
 (\ref{Illinois}) -- (\ref{MAX-lab}) and described
in \cite{levc00}, it is very disturbing that other calculations 
\cite{chen98,karakovski99,beane99,griesshammer02}
show large inconsistencies. Therefore, it appears too early to consider
the numbers  given in  (\ref{Illinois}) -- (\ref{MAX-lab}) as final.

As a summary we can make the following statement. 
Differing from the methods of electromagnetic scattering of neutrons
in a Coulomb field and coherent elastic (Compton) scattering by the deuteron,
the present method of quasi-free Compton scattering by the neutron
at energies between pion threshold and the $\Delta$ peak
is well tested and found valid. This confirmation of the method
has been achieved by carrying out the same reaction
on the proton and comparing the results for the electromagnetic 
polarizabilities obtained through this method with the ones  
obtained for the free proton.

The present publication contains an update and extension of a short
report \cite{kossert01} on first measurements of differential cross
sections for quasi-free Compton scattering by the proton and the
neutron covering a large energy interval from $E_\gamma = 200$ to
\unit[400]{MeV}. This large coverage is indispensable for determining
data for the electromagnetic polarizabilities with good precision.

\section{Experiment}
\begin{figure}
  \centering
  \includegraphics[width=0.9\columnwidth]{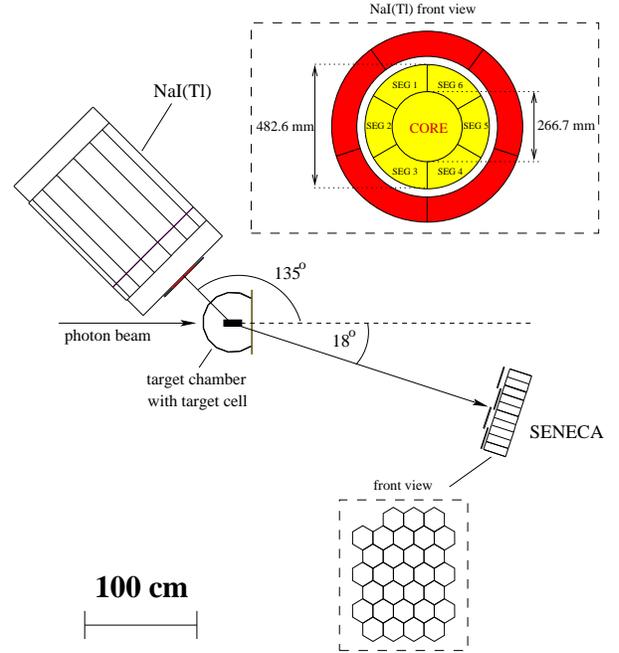}
  \caption{\label{seneca_setup}The experimental setup used to
    measure quasi-free Compton scattering from the bound neutron and
    proton.  The scattered photons were detected with the large volume
    NaI(Tl) detector, the recoiling neutrons and protons with the
    SENECA detector system. Liquid deuterium and liquid hydrogen have been used
    as target materials.  The target cells are mounted in a scattering
    chamber having a  Kapton window downstream the photon beam to
    reduce the energy loss of the protons on their way to SENECA.}
\end{figure}

The experimental setup installed at the MAMI tagged-photon facility 
\cite{anthony91}
is
outlined in Fig.~\ref{seneca_setup}.  The large Mainz \unit[48]{cm}
$\O$ $\times$ \unit[64]{cm} NaI(Tl) detector \cite{wissmann99,wiss94} was
positioned at a scattering angle of 
$\theta^{\rm LAB}_\gamma = 136.2^\circ$. The
energy resolution of this detector is \unit[1.5]{\%} in the $\Delta$
energy region and its detection efficiency \unit[100]{\%}.  The recoil
nucleons were detected with the G\"ottingen SENECA recoil detector
\cite{seneca} positioned at an emission angle of $\theta_N =
-18^\circ$, thus covering the angular range corresponding to the quasi-free
kinematics.  A distance of \unit[250]{cm} was  chosen to
compromise between a reasonable energy resolution for the
time-of-flight measurement, $\Delta E_{n}/E_{n} \approx
\unit[10]{\%}$, and the geometrical acceptance, being
$\Delta\Omega_{n}\approx \unit[90]{msr}$.  As target a \unit[5]{cm}
$\O$ $\times$ \unit[15]{cm} Kapton cell filled with liquid deuterium
was used. By filling the same cell with liquid hydrogen it was possible
to investigate quasi-free and free Compton scattering on the proton 
under exactly identical kinematical conditions.

SENECA was built as a neutron detector capable of pulse-shape
discrimination.  It consists of 30 hexagonally shaped cells 
filled with NE213 liquid scintillator  
(\unit[15]{cm} in diameter and \unit[20]{cm} in length)
mounted in a honeycomb structure.
Originally, this detector was designed for  neutrons
with energies up to \unit[50]{MeV}. Because
of the considerably higher neutron energies, in the present
experiment pulse-shape discrimination was not of an essential help.
Nevertheless, the detector had ideal properties for the present
application, mainly because of its high granularity and because the
shape of the detector modules is  favorable for computer
simulations and detection-efficiency measurements.  Veto-detectors in
front of SENECA provided the possibility to identify charged
background particles and to discriminate between neutrons and protons.
This allowed  clean separation  between quasi-free Compton scattering
and $\pi^0$ photoproduction from the proton and neutron detected in
the same experiment.  The detection efficiency of the veto-detectors
for protons was implemented in the Monte Carlo program and determined
from the free-proton experiment by making use of the large amount of
$\pi^0$ photoproduction data.  The detection efficiency
was found to be \unit[99]{\%} in close agreement
with the predictions of the Monte Carlo program.

The momenta of the recoil nucleons were  measured using the
time-of-flight (TOF) technique with the NaI(Tl) detector providing the
start signal and the SE\-NE\-CA modules providing the stop signals. For
that purpose each SENECA module had to be time calibrated relative to
the NaI(Tl) detector and this  was carried out in two steps. First, the
veto-detector in front of the NaI(Tl) detector was time calibrated
relative to the NaI(Tl) detector.  Thereafter, this veto-detector was
mounted underneath the SENECA detector so that cosmic-ray  muons could be
used for the time calibration of the SENECA modules relative to the
veto-detector. Walk corrections of the detectors were performed by
measuring time differences relative to the tagger signal as a function
of pulse height.

Data were collected during \unit[238]{h} of beam time with a deuterium
target.  In a separate run of  \unit[35]{h}  the same target cell 
was filled with
liquid hydrogen in order to measure Compton scattering from the free
proton at exactly the same kinematical conditions. These latter data 
were also used to measure
the neutron detection efficiency of SENECA
as described in subsection~\ref{sec:neut-detec-eff}.  The tagging
efficiency, being \unit[55]{\%}, was measured several times during the
runs by means of a Pb-glass detector moved into the direct photon
beam, and otherwise monitored by a P2-type ionization chamber
positioned at the end of the photon beam line.

\section{Data analysis}

\subsection{Identification of 
$(\gamma,\gamma)$ and $(\gamma,\pi^0)$ events}
Before analysing the data obtained with the deuterium  target the
corresponding analysis of data obtained with the hydrogen target was
carried out.  In this case the separation of events from Compton
scattering and $\pi^0$ photoproduction can  be achieved by the
NaI(Tl) detector alone, but the detection of the recoiling proton improves
the separation, especially for energies near the peak of the $\Delta$
resonance.  A typical spectrum is shown in panel {\bf a} of
Fig. 2. The data obtained for the free proton have been used to
optimize the analysis procedure for the bound nucleon described in the
following.

The time-of-flight  measured for protons is modified 
through energy losses in the
target and in the materials between target and the SENECA modules. The
necessary correction was  carried out through an iteration.
Starting with an estimated kinetic energy of the proton, the energy
loss was calculated in steps of a few mm in all materials traversed by
the proton. To achieve good precision in a minimum of computer time,
the step size was chosen to depend on the material.  The calculated
TOF thus achieved was compared with the measured one. In case of a
difference the estimated initial energy of the proton was modified and
the TOF was calculated again. By this procedure the initial kinetic
energy of the proton was found after few iterations.  Except for the
energy loss and a slightly different effect of the spatial resolution
of the SENECA modules, the kinetic energy of the recoiling neutrons was
determined in the same way.

\begin{figure}
  \centering
  \includegraphics[width=1.0\columnwidth]{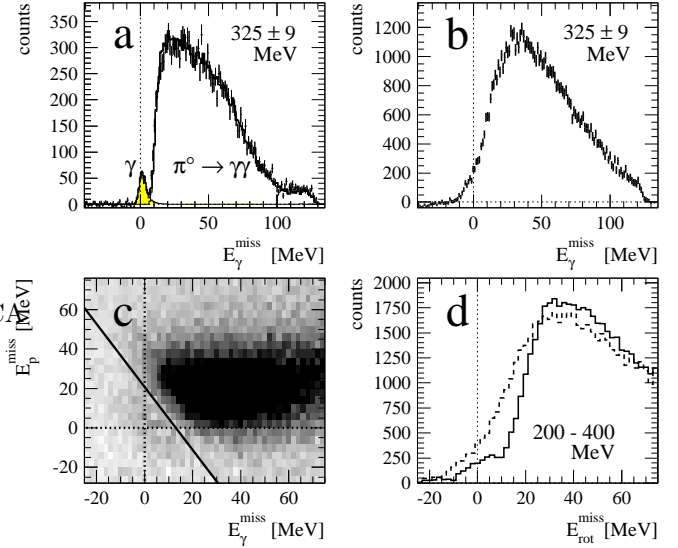}
  \caption{\label{miss_2d} Panel {\bf a}:
    Number of  proton events obtained with a hydrogen target versus
    the missing photon energy $E^{\rm miss}_{\gamma}$. Panel {\bf b}: 
    The same as panel {\bf a} but
    for proton events obtained with a deuterium  target. 
    Panel {\bf c}: Scatter plot of
    proton events obtained with a deuterium  target. Abscissa and
    ordinate are the missing energies of the scattered photon and the
    recoil proton, respectively.  The thick solid line separates
    Compton events located in the vicinity of the origin of the
    coordinate system and $(\gamma,\pi^0)$ events shown as a dark
    range. For the further evaluation each point in the scatter plot
    was rotated (moved on a circle centered at  the origin), until the
    thick solid line was perpendicular to the abscissa.
    Panel {\bf d}:
    Projection of proton events obtained with a deuteron target after
    rotation (solid line) and before rotation (broken line).}
\end{figure}
\begin{figure}
  \centering
  \includegraphics[width=1.0\columnwidth]{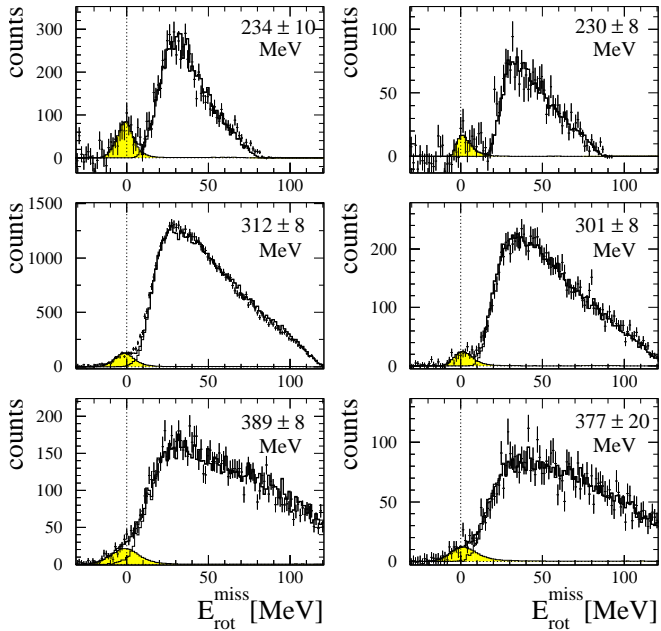}
  \caption{\label{miss_1d}Typical spectra of events obtained with
    the deuteron target shown for the recoil proton (left panels) and
    recoil neutrons (right panels). The data from the two dimensional
    plot have been projected on the abscissa after the rotation 
    described in the caption of Fig.2 and in the text has
    been carried out.  The solid curves are the results of a Monte
    Carlo simulation scaled to the Compton events and the
    $(\gamma,\pi^0)$ events, respectively.}
\end{figure}

For the separation of events from Compton scattering and $\pi^0$
photoproduction it is convenient to use two-dimensional scatter plots
of events with the missing nucleon  energy 
$E^{\rm miss}_N = E^{\rm calc}_N  - E^{\rm SEN}_N$ 
and the missing photon energy  
$E^{\rm miss}_\gamma = E^{\rm calc}_\gamma -E^{\rm NaI}_\gamma$ 
as the parameters, where $E^{\rm SEN}_N$ and $E^{\rm NaI}_\gamma$
denote measured energies and $E^{\rm calc}_N$ and $E^{\rm calc}_\gamma$
the corresponding calculated energies. The calculations are carried out
using the tagged photon energy and the detected nucleon angle or  scattered
photon angle and 
assuming the kinematics
of Compton scattering by the proton  in case of a hydrogen target or the
kinematics of Compton scattering in the center of the quasi-free peak
\cite{levchuk94}
in case of a deuteron target. As an example, the panel {\bf c} of
Fig.~\ref{miss_2d} shows the scatter plot of proton events obtained with
\unit[200]{MeV} $-$ \unit[400]{MeV}  photons  incident on a
deuterium target.  Two
separate regions containing events are visible.  The Compton events
are located in a narrow zone around the origin ($E^{\rm miss}_\gamma =
0, E^{\rm miss}_p = 0$), the $(\gamma,\pi^0)$ events in the dark range
at larger missing energies.  For the further evaluation each point in
the panel was rotated (moved on a circle centered in the origin) until
the thick solid line became perpendicular to the abscissa. This has
the advantage that projections of the data on the new abscissa 
-- denoted by $E^{\rm miss}_{\rm rot}$ --
can be used
for the further analysis without essential loss in the quality of
separation of the two types of events.

The benefits of this procedure are illustrated in 
Fig. 2.  Panel {\bf a} shows numbers of proton events 
from a proton target versus the measured
missing energy of the scattered photon, given for a narrow 
energy interval close to
the maximum of the $\Delta$-resonance. In this case we find a very
good separation of the two types of events as in previous
experiments carried out with proton targets.  The good separation
disappears when proton events of the same type are taken from a
deuteron target. These data are shown in panel {\bf b}, where it can
clearly be seen that the effects of binding destroy the separation of
the two types of events which was visible in panel {\bf a}. The separation
can be partly restored when the rotation
procedure is applied, as is shown in panel {\bf d}. This panel contains as a
solid line the same data as the scatter plot of events 
shown in panel {\bf c} but projected
on the abscissa after rotation. For comparison the broken line shows
the same data before rotation. The comparison of these two lines
clearly demonstrates that the rotation procedure essentially improves
on the separation of the two types of events. In principle it also
would have been possible to use the two dimensional scatter plot
directly for the separation. However, this
latter procedure would have been difficult because of the limited
statistical accuracy  which is even less favorable for the 
case of neutron events.

Figure~\ref{miss_1d} shows typical spectra  obtained with the
deuterium target. The left panels contain proton events, the right panels
neutron events. The different numbers of events on the two sides are
due to the neutron detection efficiency which was \unit[18]{\%}.  As
expected from the findings of panel {\bf d} of Fig.~\ref{miss_2d},
there is a reasonable separation of the two types of events in the
whole energy range. For the final separation and for the determination
of the numbers of Compton events, a complete Monte
Carlo simulation has been carried out for the processes under
consideration.  In order to achieve  high precision, all  details  of the
experiment were  implemented with great care. The results of these
simulations shown by solid curves were scaled to the Compton and
$(\gamma,\pi^0)$ data, thus leading to the grey areas in case of the
Compton scattering events and to the white areas in case of the
$\pi^0$ events.

\subsection{Details of Monte Carlo simulations and 
differential cross sections}

\subsubsection{Compton scattering 
by the free proton}

In case of the free proton (hydrogen target), the differential cross
section is given by the number of scattered photons, the incident
photon flux and the solid angle of the NaI(Tl) detector. The
determination of the latter needs a rather detailed Monte Carlo
simulation of the exact experimental  setup including target position,
beam profile and all detector materials and resolutions.  The
simulation is based on the package GEANT provided by the CERN
computing department \cite{brun93}.  The event generator uses the
angular and energy distributions for Compton scattering calculated
with the dispersion approach by L'vov {\it et al.}~\cite{LPS97}.

Since there is a large component due to
$\pi^0$ pho\-to\-pro\-duc\-tion in the missing-energy spectra, also
this process has to be included in the simulation. For free
$\pi^0$ photoproduction, the angular and energy distributions were
calculated using the SAID partial wave analysis
\cite{arndt96,said}.  

For both Compton scattering and $\pi^0$ photoproduction from the free
proton  the missing-energy spectra, like the one shown in
Fig.~\ref{miss_2d} panel {\bf a}, show  very good agreement between
the measured and simulated spectra after the scaling to the experimental data
has been carried out. Therefore, the experimental setup
is well described in our Monte Carlo program.  Comparing the measured
and  simulated spectra  gives the number of scattered photons. 
Differential cross sections are then obtained by normalizing 
to the number of target nuclei and to the incident photon flux.

\subsubsection{Quasi-free Compton scattering by the nucleon}

In case of quasi-free Compton scattering and $\pi^0$ photoproduction,
the aim of the analysis is to determine the triple differential cross
section in the center of the quasi-free peak.  For this purpose the
Monto-Carlo simulation as described above has to be extended in order
to also include the effects of the quasi-free reactions.  For
quasi-free Compton scattering the theoretical treatment of
\cite{levchuk94,wissmann98} was used for this purpose whereas quasi-free
$\pi^0$ photoproduction was based on the treatment given in
\cite{LLP96}.  The events have to be distributed along all possible
kinematical variables since they are only strongly correlated
in the center of the quasi-free peak.
After all variables were selected for a Monte Carlo event the weight
of such an event was  determined according to the theoretical
cross section. The geometrical boundary conditions set by the
acceptances of the detectors were sufficiently enlarged in order to
make sure that the final sample of Monte Carlo events was complete.
The agreement between experiment and simulation is demonstrated in
Fig.~\ref{miss_1d} and has already been discussed in the previous
section.

The number of Compton scattered events, which is the
number of events as given by the adjusted curves, corresponds to the
integral of the triple differential cross section in the region of the
quasi-free peak. 
The following relation has been used to determine the final 
triple differential cross section in the center of the
nucleon quasi-free peak (NQFP):
\begin{equation}\label{wq_qf}
  \left( \frac{d^3\sigma}{{d\Omega_\gamma}{d\Omega_N}{dE_N}}\right)_{NQFP} 
  = \frac{N_{\gamma N}}
  {N_\gamma N_T  \epsilon_N R^{\gamma\gamma}_{NQFP}}\; ,
\end{equation}
where $N_{\gamma N}$ are the number of coincident $\gamma$-nucleon
events as extracted from the missing-energy spectra, $N_{\gamma}$ are
the number of incident photons, $N_T$ are the number of target nuclei,
$\epsilon_N$ is the nucleon detection efficiency and
$R^{\gamma\gamma}_{\rm NQFP}$ is a factor obtained by a Monte Carlo
simulation which relates the number of scattered photons integrated
over the Compton peak to the triple differential cross section in the
center of the nucleon quasi-free peak.

\subsection{Neutron detection efficiency}\label{sec:neut-detec-eff}
The neutron detection efficiency can  be determined  experimentally
using neutrons from the reaction $p(\gamma,\pi^+ n)$ on the free
proton, with the $\pi^+$ meson detected by the NaI(Tl) detector and
identified as a charged particle by the veto-counter in front of it.
The NaI(Tl) detector measures the kinetic energy of the $\pi^+$ meson,
so that after the necessary  corrections for energy losses in the 
materials on the way from the production point to the NaI(Tl) 
detector are carried out, the experimental
initial kinetic energy $E_{\pi^+}^\text{NaI}$ at the reaction point 
is known. The same quantity can also be calculated from the incident
photon energy and the $\pi^+$ emission angle, leading to
$E_{\pi^+}^\text{calc}$. From these two values for the $\pi^+$ energy, the
missing $\pi^+$-energy
\begin{equation} 
  E_{\pi^+}^\text{miss} = E_{\pi^+}^\text{calc} - E_{\pi^+}^\text{NaI},
\end{equation}
can be calculated and used for the identification of a \\ $p(\gamma,\pi^+ n)$
event.
\begin{figure}
  \centering
  \includegraphics[width=1.\columnwidth]{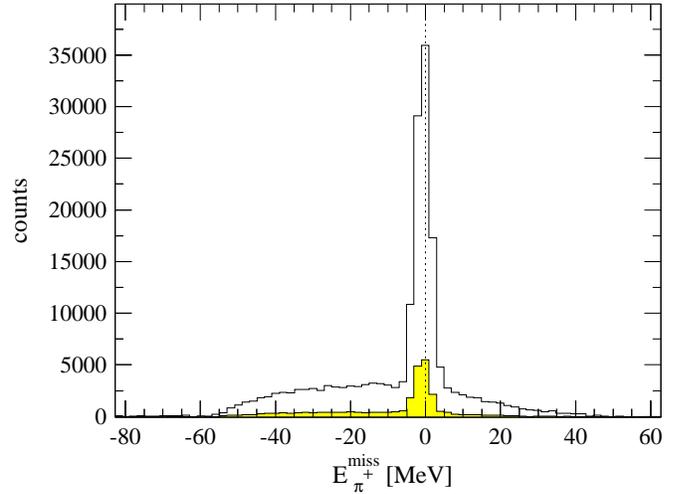}
  \caption{\label{missing_pion}Missing kinetic energy of the $\pi^+$
meson from the 
    reaction $p(\gamma,\pi^+)$ with the $\pi^0$ meson detected
    by  the NaI(Tl) detector (solid line) and the same data as obtained in
    coincidence with a neutron identified with SENECA (grey area).}
\end{figure}
The spectrum of events versus  the missing energy as shown in
Fig.~\ref{missing_pion} has  the expected structure with a narrow
peak around zero missing energy containing the cleanly identified 
$p(\gamma,\pi^+ n)$ events on top of a broad background.

\begin{figure}
  \centering
  \includegraphics[width=1.\columnwidth]{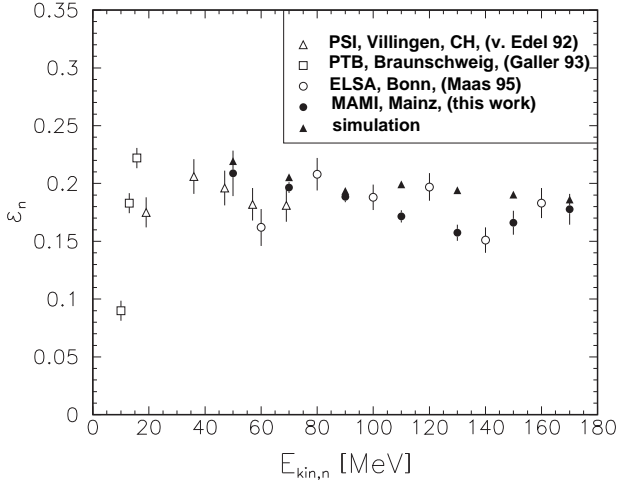}
  \caption{\label{sen_eff}The neutron detection efficiency of
    SENECA as measured in the present experiment via the reaction
    $p(\gamma,\pi^+ n)$ (full circles).}
\end{figure}

Since the kinematical quantities are  fixed 
through the $\pi^+$ meson spectrometrized by the NaI(Tl) detector,
the corresponding recoiling neutron will hit  the SENECA detector. 
Therefore, the same missing energy spectrum is obtained in coincidence 
with the SENECA detector, except for a factor which is equal to the neutron
detection efficiency $\epsilon_n$. The coincidence spectrum is shown by the
grey area in Fig.~\ref{missing_pion}. 
The result of the present  measurement is represented  in Fig.~\ref{sen_eff}
by filled circles. It is obvious that the present data have comparatively
small errors and, therefore,  largely reduce
the uncertainties in the neutron detection efficiencies $\epsilon_n$
contained in previous measurements,  obtained with
monoenergetic neutrons at the Paul-Scherrer-Institute (Villigen,
Switzerland) \cite{edel92}, at the Physikalisch-Technische
Bundesanstalt (Braunschweig, Germany) \cite{edel92,gall93} and at the
electron accelerator ELSA (Bonn, Germany) \cite{maas95}.  
For the simulation the code GCALOR \cite{zeit96} was available 
which realistically
takes into acount the interaction of neutrons and charged pions with
the materials   down to energies of $\sim$ 1 MeV. This program, however,
requires very much computer time. Therefore, the much faster 
Gamma-Hadron-Electron-Interaction SH(A)ower
code GHEISHA \cite{fese85} was used which is included in GEANT as a standard.
Uncertainties contained in this latter code were eliminated by correcting
the simulated results
with the ratio of the results of the present experiment and
the simulation. 

\section{Theory}
\subsection{Definition of electromagnetic polarizabilities}
Following \cite{lvov93} we define the electric polarizability
$\alpha$ and the magnetic polarizability $\beta$ through the
relations
\begin{eqnarray}
{\bf D}&=& 4 \pi\, \alpha\, {\bf E},\label{electric} \\
{\bf M}&=& 4 \pi\,  \beta\,  {\bf H},\label{magnetic}\\
V_{\rm pol}(\alpha,\beta)&=& - \frac12 (4 \pi \,  \alpha)\, {\bf E}^2
- \frac12 (4  \pi \, \beta)\, {\bf H}^2. \label{def}
\end{eqnarray}
In (\ref{electric}) -- (\ref{def}) ${\bf E}$ and  ${\bf H}$ are the 
electric and magnetic
fields, respectively, ${\bf D}$ and  ${\bf M}$ the induced electric
and magnetic dipole moments, respectively, and $V_{\rm pol}$ is
the polarization potential. The factors $4 \pi$ introduced
in (\ref{electric}) -- (\ref{def}) indicate that electromagnetic
polarizabilities are traditionally given in Gaussian units.
The same quantities $\alpha$ and
$\beta$ also enter into the amplitude for Compton scattering
up to the order ${\cal O}(\omega^2)$
\begin{equation}
T=T_{\rm Born} + \omega \omega' \alpha {\boldsymbol \epsilon}\cdot
 {\boldsymbol \epsilon}' + \omega \omega' \beta {\bf s}\cdot
{\bf s}' + {\cal O}(\omega^3)\label{compton}
\end{equation}
and into the differential cross section for electromagnetic
scattering of slow neutrons in the Coulomb  field of heavy
nuclei
\begin{equation}
\frac{d\sigma_{\rm pol}}{d\Omega}=
\pi M p (Ze)^2 {\rm Re} a \left\{ \alpha_n \sin\frac{\theta}{2}
- \frac{e^2 \kappa^2_n}{2 M^3} \left( 1-\sin\frac{\theta}{2}\right )\right\}.
\label{coulomb}
\end{equation}
In (\ref{compton}) $T_{\rm Born}$ denotes the Born amplitude and 
${\boldsymbol \epsilon}$, ${\boldsymbol \epsilon}'$ the unit vectors in
the directions of the electric fields of the ingoing and outgoing photons,
respectively, and   ${\bf s}$ and ${\bf s}'$ the corresponding unit vectors
for the magnetic fields. In (\ref{coulomb}) 
$p$ is  the neutron momentum
and $- a$ the amplitude for hadronic scattering by the nucleus. 
The second term in the braces is due to the
Schwinger term, i.e. the term describing neutron scattering in the Coulomb
field due to the magnetic moment only. As outlined in the introduction,
this part of the scattering
differential cross section has been omitted in previous evaluations
of neutron transmission experiments. Therefore, 
the results of those experiments have to be corrected by increasing
them by 
$e^2 \kappa^2_n/4 M^3 = 0.62$.

In some treatments the electromagnetic polarizabilities entering into
(\ref{compton}) and (\ref{coulomb}) are denoted by  $\bar \alpha$ and
$\bar \beta$ in order to descriminate them from the  quantities
entering  into (\ref{electric}) -- (\ref{def}). However, since these
quantities are identical \cite{lvov93},
there is no good reason for using different notations.

\subsection{Non-subtracted fixed-$t$ dispersion theory}

The low-energy expansion of the Compton scattering amplitude
(\ref{compton}) is valid up to photon energies of about 100 MeV.
Even with terms $\sim {\cal{O}}(\omega^4)$ included the amplitude can be used
only below pion photoproduction threshold. It has been shown experimentally
\cite{lara01,wolf01} that the fixed-$t$ dispersion theory as constructed in
\cite{LPS97} is valid up to photon energies of about 800 MeV. Therefore, an
appropriate procedure to overcome the limitations inherent in
(\ref{compton}) is to apply fixed-$t$ dispersion theory 
in its general form and to adjust its  predictions to experimental 
Compton scattering data in 
angular and energy regions where the sensitivity of the differential cross
section to  the quantity under consideration is strong.

In the general case Compton scattering is described by six invariant amplitudes
$A_i(\nu,t)$, $i=1\cdots 6$    where 
\begin{eqnarray}
\nu&=&\frac{s-u}{4m}=E_\gamma+\frac{t}{4m},\quad t=(k-k')^2,\quad s=(k+p)^2,
\nonumber\\
u&=&(k-p')^2
\label{T4}
\end{eqnarray}
and $s+u+t=2m^2$, with $m$ being the nucleon mass. These amplitudes 
can be constructed to have no kinematical
singularities and constraints and to obey the usual dispersion relations.
In \cite{LPS97} fixed-$t$ dispersion relations for $A_i(\nu,t)$ 
were formulated by using a
Cauchy loop of finite size (a closed semicircle of radius $\nu_{max}$),
so that \cite{LPS97}
\begin{equation}
{\rm Re} A_i(\nu,t)=A^{\rm pole}_i(\nu,t) 
+A^{\rm int}_i(\nu,t)+A^{\rm as}_i(\nu,t)
\label{T5}
\end{equation}
with 
\begin{eqnarray}
A^{\rm pole}_i(\nu,t) &=&\frac{a_i(t)}{\nu^2-t^2/16m^2},\nonumber\\
A^{int}_i(\nu,t)&=&\frac{2}{\pi}{\cal P}\int^{\nu_{\rm max}(t)}_{\nu_{\rm thr}
(t)} {\rm Im} A_i(\nu',t)\frac{\nu' d\nu'}{\nu'^2 - \nu^2},\nonumber\\
A^{\rm as}_i(\nu,t)&=&\frac{1}{\pi}   {\rm Im} \int_{{\cal C}_{\nu_{\rm max}}}
 A_i(\nu',t)\frac{\nu' d\nu'}{\nu'^2 -\nu^2}.
\label{T6}
\end{eqnarray}
The explicit use of the contour integral for $A^{\rm as}(\nu,t)$
is only necessary for i = 1 and 2, where special models have to be used 
for this purpose. For i = 3 $\cdots$ 6 the contour integral for 
$A^{\rm as}(\nu,t)$ can be avoided by extending the integral for
$ A^{int}(\nu,t)$ to infinity. 
  
The integral contributions $A^{\rm int}_i(\nu,t)$ are 
determined by the imaginary
part of the Compton scattering amplitude which is given by the unitarity
relation of the generic form
\begin{equation}
2 {\rm Im} T_{fi}= \sum_n (2\pi)^4\delta^4(P_n-P_i)T^*_{nf}T_{ni}.
\label{T7}
\end{equation}   
The quantities entering into the r.h.s. of (\ref{T7}) are from 
$|n\rangle =|\pi N\rangle$
and $|n\rangle =|2\pi N\rangle$ intermediate states where the 
$|n\rangle=|\pi N\rangle$ component can be
constructed from recent parametrizations (SAID \cite{arndt96,said} 
and MAID \cite{drechsel99,maid}) of pion 
photoproduction  multipoles $E_{l\pm}$, $M_{l\pm}$. The  
$|n\rangle =|2\pi N\rangle$
component requires additional model considerations \cite{LPS97}.

For the asymptotic ($t-$channel) part of the amplitude $A_2(\nu,t)$ we 
may use the Low
amplitude of  $\pi^0$ exchange in the $t-$channel, supplemented by $\eta$ and 
$\eta'$ exchanges:
\begin{equation}
A^{\rm as}_2(t)= A^{\pi^0}_2(t)+A^{\eta}_2(t)+A^{\eta'}_2(t)
\label{T8}
\end{equation}
with
\begin{equation}
 A^{\pi^0}_2(t)=\frac{g_{\pi NN}F_{\pi^0 \gamma\gamma}}
{t-m^2_{\pi^0}}\tau _3 F_\pi (t),
\label{T8'}
\end{equation}
where the isospin factor is $\tau_3=\pm 1$ for the proton and neutron,
respectively, and the product of the $\pi NN$ and $\pi^0 \gamma\gamma$
couplings is \cite{LPS97}
\begin{eqnarray}
g_{\pi NN}F_{\pi^0 \gamma\gamma}&=&-16 \pi\sqrt{\frac{g^2_{\pi NN}}{4\pi}
\frac{\Gamma_{\pi^0\to 2\gamma}}{m^3_{\pi^0}}}\nonumber\\
&=&(-0.331\pm 0.012)\,{\rm GeV}^{-1}.
\label{T9}
\end{eqnarray}
The comparatively small contributions due to the $\eta$ and $\eta'$
mesons are constructed in analogy to the term representing the $\pi^0$
contribution.

In the frame of the invariant amplitudes, the backward spin polarizability
$\gamma_\pi$ is given by
\begin{equation}
\gamma_\pi= -\frac{1}{2\pi  m} \left[ A^{\rm int}_2(0,0) + A^{\rm as}_2(0,0)
+ A^{\rm int}_5(0,0)  \right]
\label{T10}
\end{equation}
with the integral ($s-$channel) part being the smaller  contribution.

There have been arguments \cite{tonnison98,blanp01}
that $A^{\rm as}_2(t)$ is not exhausted by $\pi^0$,
$\eta$ and $\eta'$
exchanges in the $t-$channel. In order to introduce an 
additional parameter into
the relevant amplitude which provides the necessary flexibility for an 
experimental test, we use the replacement
\begin{equation}
A^{\rm as}_2(t)\to A^{\rm as}_2(t)- 2\pi m \frac{\delta_{\gamma_\pi}}{
1-\frac{t}{\Lambda^2}}
\label{corrterm}
\end{equation}
from which the substitution follows:
\begin{equation}
\gamma_\pi \rightarrow \gamma_\pi +\delta \gamma_\pi.
\label{correction}
\end{equation}
The cutoff parameter $\Lambda$ of the monopole form factor
defines the slope of the function at
$t = 0$ and is chosen to be $\Lambda = 700$ MeV, as estimated from the axial
radius of the nucleon and the size of the pion \cite{LPS97}. In varying 
$\delta\gamma_\pi$ the influence of any deviation from the standard
value of $\gamma_\pi$ can be investigated in terms of this ansatz.
It  should be noted that  this procedure is completely decoupled from the
choice  of the parametrization of the photo-meson amplitudes 
$E_{l\pm}$ and $M_{l\pm}$.

Instead of calculating the $s-$channel part of $\gamma_\pi$ from the integral
amplitudes $A^{\rm int}_2(0,0)$ and $A^{\rm int}_5(0,0)$ of (\ref{T10})
the sum rule derived in \cite{lvov99} may be used for this purpose:
\begin{eqnarray}
\gamma_\pi(s-{\rm{channel}})&=
\int^\infty_{\omega_0} \sqrt{1+\frac{2 \omega}{m}}\left( 1+\frac{\omega}{m}
\right)\nonumber\\&\times \sum_n P_n [\sigma^n_{3/2}-\sigma^n_{1/2}]
\frac{d\omega}{4 \pi^2 \omega^3}
\label{sumrule}
\end{eqnarray}
where $\sigma^n_{\lambda}$ is the photoabsorption cross section with 
specific total helicity $\lambda$ of the beam and target and with
relative parity $P_n=\pm 1$ of the final state $n$ with respect to the target.

The asymptotic contribution of the amplitude $A_1(\nu,t)$ is modeled
through an ansatz analogous  to the Low amplitude, except for the fact
that the pseudoscalar meson $\pi^0$ is replaced by the  scalar
$\sigma$-meson. 
In this case we use a simpler form  of the ansatz
\begin{equation}
A^{\rm as}_1(t)\simeq A^\sigma_1(t)=\frac{g_{\sigma NN}F_{\sigma\gamma\gamma}}
{t-m^2_\sigma}
\label{T10'}
\end{equation}
and include quantities like the formfactor in (\ref{T8'}) into the 
"effective mass" $m_\sigma$ being  now  an adjustable 
parameter \cite{LPS97}. 
The quantity
$g_{\sigma NN}F_{\sigma \gamma\gamma}$ is given by the difference of 
the electric and magnetic polarizabilities $\alpha--\beta$ through
\begin{equation}
2 \pi(\alpha-\beta)+A^{\rm int}_1(0,0)= -A^{\rm as}_1(0,0)=
\frac{g_{\sigma NN} F_{\sigma\gamma\gamma}}{m^2_\sigma}
\end{equation} 
with the integral part being a minor contribution.
The quantity $m_\sigma$ has a strong impact on the differential cross section
in the 400 to 700 MeV energy region and has been determined
experimentally to be $m_\sigma \approx 600$ MeV \cite{lara01,wolf01}. 

For the proton $\alpha_p -\beta_p$ has been  determined in 
a series of Compton scattering 
experiments carried out below $\pi$ photoproduction threshold.
The results may be summarized by quoting  the  global average 
\cite{olmos01}
$\alpha_p -\beta_p
=10.5\pm 0.9({\rm stat+syst}) \pm 0.7({\rm model})$. Adopting this value,  
the correction $\delta \gamma_\pi$ introduced in (\ref{correction})
has been determined to be comparable  with $\delta \gamma^{(p)}_\pi = 0$
\cite{camen02}.

In the present experiment $\alpha_n - \beta_n$ will be determined 
from experimental differential cross sections measured at energies
above $\pi$ threshold. This strongly reduces the possibility to disentangle the
determination of $\alpha_n - \beta_n$ and 
$\gamma^{(n)}_\pi$. Our procedure, therefore,
is to assume that the result  $\delta \gamma_\pi = 0$ obtained for the 
proton  is also true
for the neutron and then to extract $\alpha_n - \beta_n$ 
from fits to the experimental data. This point will be discussed
in more detail later.

\subsection{Quasi-free Compton scattering}
\begin{figure}
  \centering
  \includegraphics[width=1.\columnwidth]{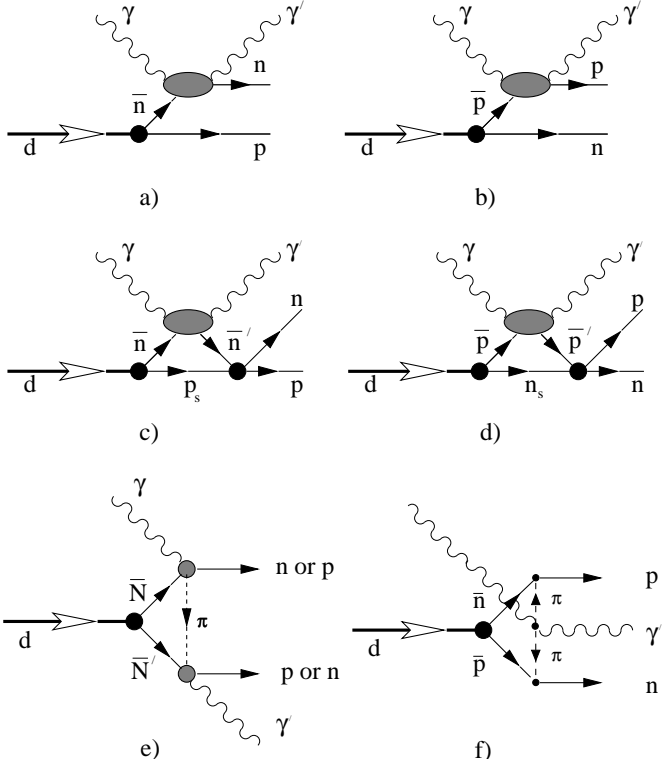}
  \caption{\label{graphs}Main graphs contributing to the
    reaction $\gamma d\to \gamma^\prime np$. The shaded areas in graph
    e) stand for the $\gamma N\to \pi N$ and $\pi N\to \gamma N$
    amplitudes including pole and $\Delta$ terms.  }
\end{figure}

A calculation of the reaction $\gamma d\to \gamma^\prime np$ has been
carried out in Ref.~\cite{levchuk94} (see also Ref.~\cite{wissmann98})
in the framework of a diagrammatic approach. The main graphs
contributing to the reaction amplitude in the quasi-free region are
displayed in Fig.~\ref{graphs}. Graphs \ref{graphs}a)
and b) describe quasi-free scattering from the neutron and
proton, respectively. The non-interacting nucleon in these graphs is
often referred to as a spectator. Nucleon-nucleon rescattering (or
final state interaction, FSI) shown in graphs \ref{graphs}c) and
d), as well as meson seagull contributions and isobar
configurations described by graphs \ref{graphs}e) and f)
have also been taken into account.

All the details of the calculations can be found in
Ref.~\cite{levchuk94}.  Here we mention only the following. A model
of $NN$-interaction is needed to calculate the deuteron wave function
and the $NN$-scattering amplitude when evaluating the diagrams. We
checked three versions of the Bonn OBEPR model \cite{OBEPR1,OBEPR2},
CD-Bonn potential \cite{CD-Bonn}, and a separable approximation
\cite{PEST1,PEST2} of the Paris potential \cite{Paris}.  Our
observation is that the results obtained with all the Bonn potentials
are practically the same so that in the following we will present our
results with the CD-Bonn and Paris models only.  One more important
ingredient of the model is the nucleon Compton scattering amplitude.
It enters the graphs in Fig.~\ref{graphs}a)-d). This amplitude is taken
from the currently accepted dispersion model \cite{LPS97} which
provides a satisfactory description of all available data on free proton
Compton scattering up to 800 MeV \cite{lara01,wolf01}.
As an input of the model the amplitudes of single pion photoproduction
are used. In the present work they are taken from the 
SAID \cite{arndt96,said}
(solutions SM99K and SM00K) and MAID \cite{drechsel99,maid} 
(solution MAID2000)
analyses.  Of course when considering the diagrams \ref{graphs}a)-d)
we have to  deal with off-shell nucleons. The off-shell corrections,
however, are expected to be very small in the kinematic conditions
under consideration (see a discussion in Ref.~\cite{levchuk94}) so that
the use of the on-shell parametrization for the nucleon Compton
scattering amplitudes is quite justified.

The measured triple differential cross section,\\ $d^3\sigma /
d\Omega_{\gamma^\prime}d\Omega_p dE_p $ (here $E_p$ and $\Omega_p$ are
the kinetic energy and solid angle of the proton), in the center of
the proton quasi-free peak (pQFP) can be related to the free
differential cross section, $d\sigma / d\Omega_{\gamma^\prime}$, via a
spectator formula \cite{levchuk94,wissmann99}
\begin{eqnarray}
  \frac{ d\sigma(\gamma p\rightarrow \gamma^\prime p) }
  { d\Omega_{\gamma^\prime} } =  
  \frac{ (2\pi)^3 }{ u^2(0) }
  \frac{ E_\gamma E_{\gamma^\prime} }
  { |{\bf p}|m{E_{\gamma^\prime}^{(p)2}} }
  \frac{ d^3\sigma(\gamma d\rightarrow \gamma^\prime p n) }
  {d\Omega_{\gamma^\prime}d\Omega_p dE_p },
  \label{spectator}
\end{eqnarray}
where $u(0)$ is the S-wave amplitude of the deuteron wave function at
zero momentum (the D-wave component of this function does
not contribute at zero momentum), $E_{\gamma^\prime}^{(p)}=(p_{\gamma
  ^\prime}\!\!\cdot p_p)/m$ is the final photon energy in the rest
frame of the final proton and ${\bf p}$ is the momentum of the final
proton. In case of neutron quasi-free
scattering, one has an analogous formula with the replacement 
$p\leftrightarrow n$.  Note that there is a slight dependence 
of the factor\\
 $(2\pi)^3 /
u^2(0)$ in the r.h.s.\ of Eq.~(\ref{spectator}) on the model of
$NN$-interaction. For instance, this  factor  is equal to
$\unit[592055]{MeV^3}$ and $\unit[599089]{MeV^3}$ for the CD-Bonn and for the
Paris potential, respectively, \textit{i.e.}, it varies within
\unit[1.2]{\%}.

Equation~(\ref{spectator}) is valid for the pole diagram contribution
only. Therefore, for the measured cross section to be used in
Eq.~(\ref{spectator}) the r.h.s. of this equation has to be
multiplied by a factor\footnote{In
  general, the factor $f$ depends not only on $E_\gamma$ and
  $\theta_\gamma$ but also on nucleon momenta. In the center of NQFP,
  however, one has the kinematics of a $2\to 2$ process.}
$f(E_\gamma,\theta_\gamma)=d^3\sigma_{pol}/d^3\sigma_{tot}$. Here,
$d^3\sigma_{pol}$ stands for the contribution of the pole (proton or
neutron) diagram to the total differential cross section
$d^3\sigma_{tot}$ for which all the diagrams a)-f) have been taken
into account. The difference $1-f$ shows the relative contribution of
the background effects which are mainly from FSI and ranges in the
center of the pQFP at $\theta^{\rm LAB}_\gamma=136.2^\circ$ from $-0.049$ at
\unit[234]{MeV} to $-0.014$ at \unit[389]{MeV} for the CD-Bonn
potential and from $-0.054$ to $-0.024$ for the Paris one.  In the
center of the neutron quasi-free peak (nQFP) at the same angle this
factor varies  from $-0.081$ at \unit[211]{MeV} to $-0.013$ at
\unit[377]{MeV} for the CD-Bonn potential and from $-0.097$ to
$-0.019$ for the Paris potential. The small value  of the factor $1-f$ shows
that  the background contributions due to FSI
and seagull terms are small at  the relatively high photon
energies and large  photon angle under consideration.
At energies below pion photoproduction threshold and/or for forward photon
scattering the above contributions (mainly due to FSI) are much larger
(see Refs.~\cite{levchuk94,wissmann98}).

The total factor relating the free and quasi-free cross sections
ranges from 1.48 (1.52) at \unit[210]{MeV} to 0.68 (0.70) at
\unit[380]{MeV} for CD-Bonn (Paris) potential at
$\theta^{\rm LAB}_\gamma=136.2^\circ$, \textit{i.e.}  it has only a minor
dependence on the choice of a model of $NN$-interaction.  It is seen
that the factor can lead to both lowering and increasing the "free"
cross section in comparison with the quasi-free cross section.

When relating the free and quasi-free cross section the effect of the
deuteron binding has to be taken into account because of the rather
strong energy dependence of the differential cross section. This means
that the equivalent free photon energy $E^f_\gamma$ is not equal to
$E_\gamma$ but related with it in the center of NQFP through the
equation
\begin{eqnarray}
  E^f_\gamma=E_\gamma -\Delta \left( 1 + \frac {E_\gamma - \Delta/2}{m} \right),
  \label{energy}
\end{eqnarray}
where $\Delta = \unit[2.2246]{MeV}$ is the deuteron binding energy. 
Eq. (\ref{energy}) has a sizable
effect on the construction of "free"-nucleon differential cross sections
from quasi-free triple differential cross sections.

\section{Discussion of the Results}

\begin{figure}
  \centering
  \includegraphics[width=1.\columnwidth]{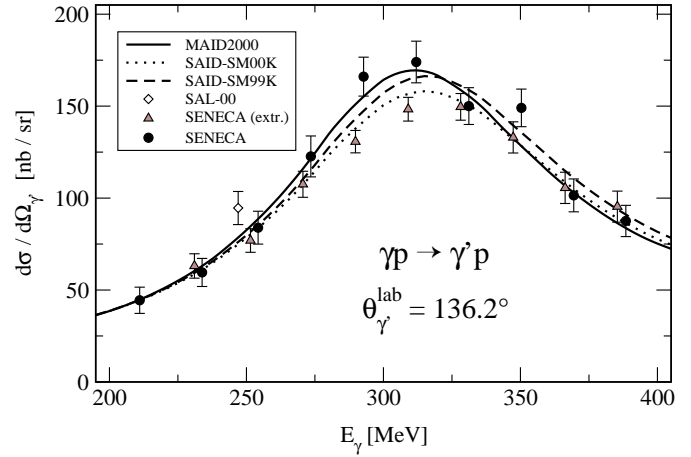}
  \caption{\label{free_p}Differential cross sections for Compton scattering
   measured on a free proton
    (filled circles) and the corresponding data  extracted from 
    quasi-free cross sections obtained 
    for the bound proton (grey triangles) at $\theta^{\rm LAB}_{\gamma} =
    136.2^\circ$.  
The SAL result \protect\cite{kolb00} is shown by a
    diamond.  Only statistical errors are shown. The curves were  obtained 
    using  the model \cite{LPS97} for different parametrizations of photomeson
    amplitudes. In all curves $\alpha_p-\beta_p=10.5$ and
    $\gamma^{(p)}_\pi=-38.7$ was used.  }
\end{figure}

The filled circles shown in Fig.~\ref{free_p} represent the experimental
differential cross sections obtained in the present experiment for
Compton scattering by the free proton. These data are compared with
predictions of the standard dispersion theory \cite{LPS97} based on different
recent parametrizations of photomeson amplitudes. 
The curves have been obtained with a 
fixed value for  the difference of the proton polarizabilities,
$\alpha_p-\beta_p=10.5$~\cite{olmos01}. Furthermore, the
standard value for the proton backward spin polarizability 
$\gamma^{(p)}_\pi=-38.7$ (in units of $10^{-4}{\rm fm}^4$ which will 
be used throughout in the following) is used
without the additional contribution introduced by
\cite{tonnison98,blanp01} for reasons discussed in \cite{camen02}.
As observed before
in a different recent experiment \cite{lara01,wolf01} there is  good 
agreement between experiment and the predictions based on the  
MAID2000  and SAID-SM99K parametrizations, whereas
the predictions based on the parametrization SAID-SM00K are too low
in the maximum of the $\Delta$-resonance. A closer look provides us with some
slight preference for the MAID2000 parametrization over the SAID-SM99K
parametrization. Therefore, we  base the
further analysis on the MAID2000 pa\-ra\-me\-tri\-za\-tion and  use the 
SAID-SM99K parametrization for estimates of model errors only. 
For comparison Fig.~\ref{free_p} also contains the corresponding 
data extracted from
quasi-free Compton scattering through the model described in section 4 with
(\ref{spectator}) being the relevant formula.
The results of this extraction procedure 
depend slightly on the choice
of the $NN$-potential but are  practically independent of the choice of
the photomeson amplitude. Average values over the CD-Bonn and Paris 
models are used here.
The agreement of the two types of data in Fig.~\ref{free_p} is good
except for the two data points at 290 and 310 MeV.
We consider this quality of  agreement as 
a satisfactory  justification for using the theory in the present form for   
the analysis of the quasi-free Compton scattering data obtained for the
neutron with the residual deviations taken into account as a contribution to
the model error.

\begin{figure}
  \centering
  \includegraphics[width=1.\columnwidth]{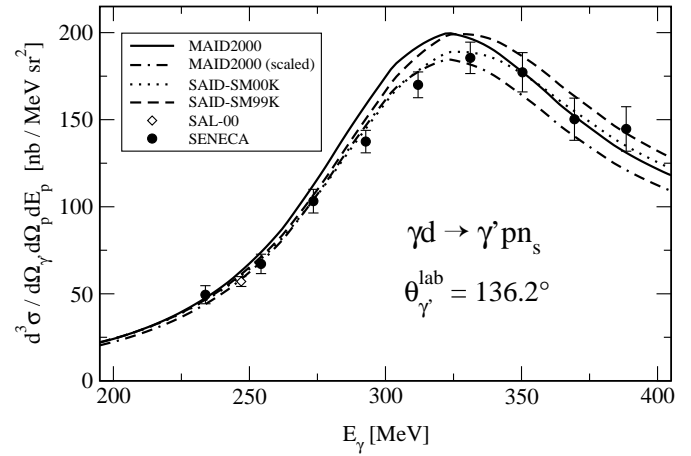}
  \caption{\label{q_free_p}Triple differential cross section of the reaction
    $d(\gamma,\gamma^{\,\prime}p)n$ in the center of pQFP at
    $\theta^{\rm LAB}_{\gamma}= 136.2^{\circ}$, given in the LAB frame.  
Circles: the present
    experiment; diamond: the SAL experiment \protect\cite{kolb00}.
    Only statistical errors are shown.  The curves were  obtained using the
    model \cite{levchuk94} for different parametrizations of photomeson
    amplitudes. In all curves $\alpha_p-\beta_p=10.5$ and
    $\gamma^{(p)}_\pi=-38.7$ was used.  }
\end{figure}

Figure~\ref{q_free_p} shows
the triple differential cross section for quasi-free Compton scattering. 
There is 
consistency between our data and the data point of the SAL
experiment at 247 MeV.
As expected, the triple  differential cross sections of Fig.~\ref{q_free_p}
reveal the same properties as the corresponding "free"  data shown 
in Fig.~\ref{free_p}.
The parametrizations MAID2000 and SAID-SM99K which have proven optimum
for reproducing the free differential cross sections show some deviations here.
At present we do not have an explanation for
this deviation.  Consequently, we have to treat it as a possible
source of uncertainty which has to be taken into account through a
further contribution to the model error. In order to get a
quantitative result for this additional model error, the prediction
shown in Fig.~\ref{q_free_p} as a solid curve has been scaled down by
a factor of 0.93 to give the dashed-dotted curve. Through this
procedure we arrive at a modified theoretical model which may be used
in the further analysis to estimate the additional model error
connected with a possible imperfection of the theory of  quasi-free
Compton scattering.

\begin{figure}
  \centering
  \includegraphics[width=1.\columnwidth]{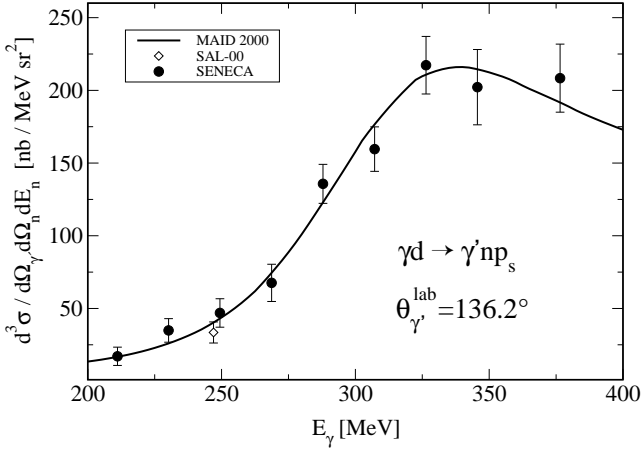}
  \caption{\label{q_free_n}
    Triple differential cross section of the reaction
    $d(\gamma,\gamma^{\,\prime}n)p$ in the center of nQFP at
    $\theta^{\rm LAB}_{\gamma}= 136.2^\circ$.
    Filled circles: the present
    experiment; diamond: SAL experiment \protect\cite{kolb00}.  Only
    statistical errors are shown.  The curve was  obtained in the
    model \protect\cite{levchuk94} with $\gamma^{(n)}_\pi = 58.6$ fixed
    and  $\alpha_n - \beta_n = 9.8$
    determined through the  fitting procedure
    as quoted in  Eq. (\ref{an-bn}).} 
\end{figure}

The measured differential cross sections for  quasi-free  Compton
scattering are displayed in Fig.~\ref{q_free_n}.  Again,  there is  consistency
with the SAL value at \unit[247]{MeV}.
Using the MAID2000  para\-me\-tri\-za\-tion of photomeson amplitudes 
which proved to be the most favorable one in case of the free proton,
we determine the polarizability 
difference $\alpha_n-\beta_n$ through a least-square procedure.
The backward spin polarizability $\gamma^{(n)}_\pi = 58.6$ of the
neutron was used as  contained in the model \cite{LPS97} for the 
MAID 2000 parametrization.
This is justified because no indication for a deviation was found in case of
the proton, i.e. $\delta^{(p)}_\pi = 0$.
The result obtained is $\alpha_n - \beta_n = 9.8$ with a reduced
$\chi^2$ of 0.6.  The errors of this result are as follows.  The
statistical error from the $\chi^2$ procedure is $\pm 3.6$.  The
systematic error of the neutron triple differential cross sections
amounts to $\pm \unit[9]{\%}$, with the detection efficiency
$\epsilon_n$ of the neutrons contributing $\pm \unit[8]{\%}$, the
number of target nuclei per cm$^2$ contributing $\pm \unit[2]{\%}$,
the uncertainties caused by cuts in the spectra and by the Monte Carlo
simulations contributing $\pm \unit[3]{\%}$ and the tagging efficiency
contributing $\pm \unit[2.5]{\%}$. For $\alpha_n - \beta_n$ this leads
to a combined systematic error of ${}^{+2.1}_{-1.1}$. The model error
due to imperfections of the parametrization of photomeson amplitudes
was estimated from a comparison of results obtained with the MAID2000
and SAID-SM99K parametrizations.  The result obtained
for $\alpha_n - \beta_n$ is $\pm 2.0$.  The errors due to different
parametrizations of $NN$-interaction were found to be about $\pm
0.2$.  The determination of the model error due to possible
imperfections of the theory of quasi-free Compton scattering amounts
to $\pm 0.8$.

\begin{figure}
  \centering
  \includegraphics[width=1.\columnwidth]{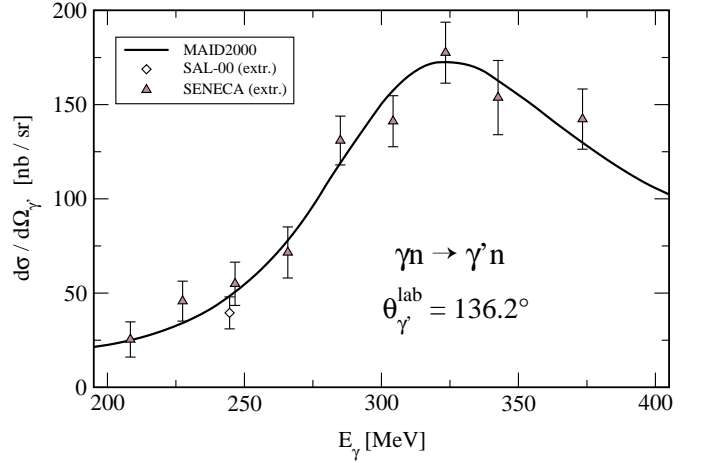}
  \caption{\label{free_n}
    Differential cross sections on the "free" neutron extracted from the
    quasi-free cross sections for the bound neutron at
    $\theta^{\rm LAB}_{\gamma}=136.2^\circ$ (gray triangles). The SAL result
    \protect\cite{kolb00} is shown by a diamond.  Only statistical
    errors are given. The curve was  obtained  in the model 
    \protect\cite{LPS97}
    using the same parameters as in  Fig. \ref{q_free_n}.
    }
\end{figure}

Taking all these errors into account we obtain  our final result
\begin{equation}
  \alpha_n - \beta_n = 9.8 \pm 3.6 \text{(stat)}
  \, {}^{+2.1}_{-1.1} \text{(syst)} \pm 2.2 \text{(model)}.
\label{an-bn}
\end{equation}
Combining it with $\alpha_n + \beta_n = 15.2 \pm 0.5$ \cite{levc00} we
obtain
\begin{eqnarray}
  \alpha_n &= 12.5 \pm 1.8 \text{(stat)} \, {}^{+1.1}_{-0.6} \text{(syst)}
  \pm 1.1 \text{(model)},\label{aln} \\
  \beta_n &= 2.7 \mp 1.8 \text{(stat)}
  \, {}^{+0.6}_{-1.1} \text{(syst)} \mp 1.1 \text{(model)}.\label{btn}
\end{eqnarray}
For sake of completeness we present in  Fig.~\ref{free_n} the
"free" cross section obtained from the quasi-free values using  the
method described above for the proton case. The consistency with the
dispersion calculation \cite{LPS97} is seen.

\begin{figure}
\centering
\includegraphics[width=1.\columnwidth]{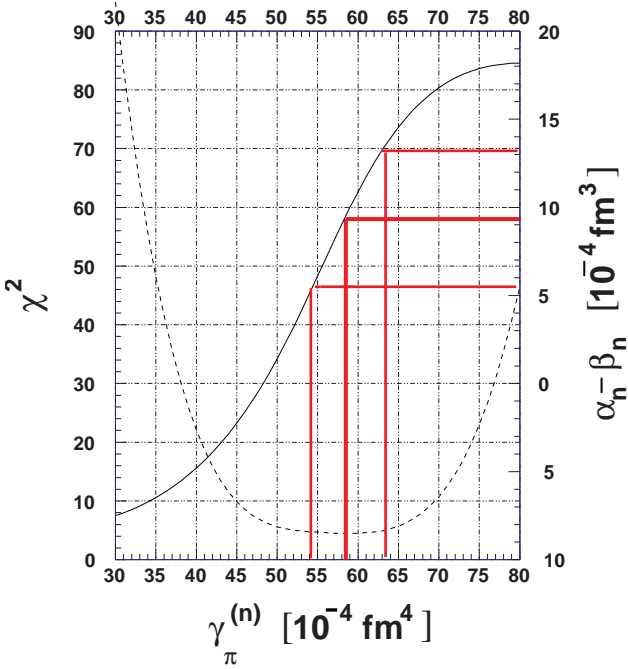}
\caption{\label{gammapi}
Solid curve: Pairs of values for $\alpha_n - \beta_n$ and $\gamma^{(n)}_\pi$ 
leading  to relative best fits to the experimental triple differential 
cross sections for quasi-free Compton scattering by the neutron. 
The corresponding $\chi^2$ is depicted by the dashed line. The 
center horizontal and vertical bars correspond to the  best fit
obtained for $\gamma^{(n)}_\pi=58.6$. 
The outer horizontal and vertical bars correspond to the statistical 
errors to be 
attributed to the  best fit.}
\end{figure}

It is of interest to compare the present results obtained for the neutron with
the corresponding result for the proton. By combining the global averages of
the electric and magnetic polarizabilities determined in Ref. \cite{olmos01}
with the value for the sum of polarizabilities $\alpha_p+\beta_p =
14.0 \pm 0.3$ obtained in \cite{levc00}, we arrive at
\begin{eqnarray}
&&\alpha_p=12.2 \pm 0.3({\rm stat})\mp 0.4({\rm syst})\pm 0.3({\rm model}),\\
&&\beta_p=1.8 \pm 0.4({\rm stat})\pm 0.4({\rm syst})\pm 0.4({\rm model}).
\end{eqnarray} 

The backward spin polarizability for the neutron as provided by the 
MAID2000 parametrization through (\ref{T10}) 
is $\gamma^{(n)}_\pi= 58.6$.
For comparisons with theoretical predictions it is of interest to get
some experimental information on this number. This
is investigated in Fig. \ref{gammapi}
showing the result of a least-square procedure of the same
kind as used for the determination of
$\alpha_n - \beta_n$.  The only  difference is that  
$\gamma^{(n)}_\pi$ was not kept constant at 
$\gamma^{(n)}_\pi= 58.6$
but  both quantities   $\alpha_n - \beta_n$ and $\gamma^{(n)}_\pi$
were treated as free parameters. Then for each adopted value of
$\gamma^{(n)}_\pi$ a corresponding value for  $\alpha_n - \beta_n$
was obtained, with  $\chi^2$ being at its minimum. The pairs
of values for the two quantities $\gamma_\pi$  and  $\alpha_n - \beta_n$ 
shown on the abscissa and the right ordinate, respectively, are 
depicted by a solid curve, the corresponding minimum of $\chi^2$ 
by a dashed curve. The center horizontal
and vertical bars depict those pairs of $\alpha_n - \beta_n$
and  $\gamma^{(n)}_\pi$ which correspond to the best fit
to the experimental triple differential cross sections for quasi-free
Compton scattering by the neutron with the spin polarizability
still kept  to be
$\gamma^{(n)}_\pi=58.6$. 
Though the $\chi^2$ distribution in Figure \ref{gammapi} is  rather 
broad, some information on $\gamma^{(n)}_\pi$ may be obtained
going beyond the assumption that this quantity should be identical to the
number resulting from  Eq. (\ref{T10}). Apparently, $\gamma^{(n)}_\pi = 58.6$
is very close to the center of the $\chi^2$ distribution and in this sense
may be considered as being confirmed by the experiment. 
In order to arrive at some 
measure for  the error of  $\gamma^{(n)}_\pi$, we introduce in Fig.
\ref{gammapi} the statistical error of $\alpha_n - \beta_n$ as
outer horizontal bars and take the "statistical" 
error of $\gamma^{(n)}_\pi$ from the corresponding outer vertical bars.
This leads to the final result 
\begin{equation}
\gamma^{(n)}_\pi= 58.6 \pm 4.0.
\label{finalresult}
\end{equation}

Table \ref{table} summarizes the results obtained for the electromagnetic
structure constants of proton and neutron in recent Compton
scattering experiments. The experimental backward-angle spin polarizabilities
are split up into  $t-$channel and  $s-$channel parts. 
\begin{table}
\caption{Structure constants for the proton and the neutron.
The quantities in lines a -- c given for the neutron are from the present work,
those given for the proton from our recent works \cite{olmos01} and 
\cite{camen02}. The numbers in lines d and  f are from \cite{lvov99}.
The numbers in line e are calculated from lines c and d.
}
\label{table}
\begin{center}
\begin{tabular}{llll}
\hline
structure constant&proton&neutron\\
\hline
$\alpha$&$12.2 \pm 0.6$ & $12.5\pm 2.3$&a\\
$\beta$&$ 1.8\mp 0.6$& $2.7\mp 2.3$&b\\
$\gamma_\pi $&$-38.7\pm 1.8$&$+58.6\pm 4.0$&c\\
$\gamma_\pi(t-{\rm{channel}})$&$-46.6$&$+43.4$&d\\
$\gamma_\pi( s-{\rm{channel}})$&$+7.9\pm 1.8$&$+15.2\pm 4.0$&e\\
\hline
$\gamma_\pi(s-{\rm{channel}})$&$+7.1\pm 1.8$&$+9.1\pm 1.8$&f\\
\hline
\end{tabular}
\end{center}
\end{table}
The $t-$channel contribution is given by the Low amplitudes of Eqs. 
(\ref{T8}) and (\ref{T10}), the $s-$channel contribution by the integral
amplitudes $A^{\rm int}_2(0,0)$ and $A^{\rm int}_5(0,0)$ of Eq. (\ref{T10}).
It is interesting to note that the $s-$channel contribution has a strong
isovector part. This isovector part is smaller
when using the sum rule of Eq. (\ref{sumrule}) for the determination
of these quantities \cite{lvov99}. This simultaneously leads to a 
total $\gamma^{(n)}_\pi$ = 52.5 $\pm$ 2.4 \cite{lvov99} which hardly 
may be considered as consistent with the 
present experiment as seen in Figure \ref{gammapi}. We tentatively
assume that the large uncertainties in the photomeson
amplitudes still existing in the second resonance region are 
responsible for this discrepancy.

\begin{table}
\caption{Summary on  neutron polarizabilities. Lines 2 -- 4 contain the
results obtained by electromagnetic scattering of neutrons
in a Coulomb field \cite{schmiedmayer91}, 
by the method of quasi-free Compton scattering above $\pi$ threshold 
(present work) and
by coherent Compton scattering off the deuteron \cite{lundin02}. }
\label{global}
\begin{center}
\begin{tabular}{lll}
\hline
$\alpha_n$+$\beta_n$&15.2$\pm$ 0.5& Baldin's sum rule \cite{levc00}\\
\hline
$\alpha_n$& 12.6$\pm$ 2.5&Coulomb field \cite{schmiedmayer91}\\
$\alpha_n$& 12.5$\pm$ 2.3&quasi-free Compton (present work)\\
$\alpha_n$&  9.2$\pm$ 2.2&deuteron coherent \cite{lundin02}\\
\hline
\end{tabular}
\end{center}
\end{table}

Three independent methods  for determining the electric  polarizability
of the neutron exist,
viz. the electromagnetic scattering of neutrons in the Coulomb field
of heavy nuclei, the quasi-free Compton scattering off neutrons bound in the
deuteron and coherent Compton scattering by the deuteron. Table \ref{global}
contains one  representative result for each method. In case of
electromagnetic scattering of neutrons in a Coulomb field the result was
selected where the highest precision was claimed by the authors.
For coherent Compton scattering by the deuteron
the most recent  experiment has been selected.
It appears that the three results are of comparable
precision and are in agreement with each other within a 1-$\sigma$ error.
However, a closer look reveals that this agreement may be fortuitous because of
large uncertainties possibly contained in the results of
lines 2 and 4, as discussed above. Therefore, only the present experiment
where all possible errors have been analysed with great care
may be considered as reliable.

The obtained value for $\alpha_n$ is consistent with ChPT prediction
at ${\cal O}(q^4)$ \cite{BKM95}, $\alpha_n =13.4\pm 1.5$, but for
$\beta_n$ one has a noticeable disagreement, $\beta_n=7.8\pm 3.6$. The
most recent results in the framework of heavy baryon ChPT with the
$\Delta$ isobar included
\cite{hemmert98}, $\alpha_n =16.4$ and $\beta_n =9.1$, contradict
both measured  values (\ref{aln}) and (\ref{btn}). Recently a covariant
"dressed $K$-matrix model" has been built~\cite{kondr01} and its
predictions for the dipole and spin polarizabilities of the nucleon
have been given. The results for the proton are  
$\alpha_p = 12.1$, $\beta_p = 2.4$
and $\gamma^{(p)}_\pi (s-{\rm{channel}})= 11.4$, and for the neutron 
$\alpha_n = 12.7$, $\beta_n = 1.8$
and $\gamma^{(p)}_\pi(s-{\rm{channel}})= 11.2$. The results for 
$\alpha$, $\beta$ and $\gamma_\pi$ appear to be in reasonable agreement
with the numbers given in Table 1, except for the fact that the predicted
$\gamma_\pi$ has no isovector component.

The experimental data of the present experiment shown in Figs. 7 -- 10
are also given in tabular form in the appendix.
\section{Conclusion}

The results of the present experiment can be summarized as follows.
The energy dependence of the differential cross section for  Compton
scattering off the quasi-free proton and, for first time, off the
quasi-free neutron has been measured in the energy region 
from 200 to \unit[400]{MeV}. In agreement with the corresponding result for
the proton, it is shown that the backward spin polarizability 
$\gamma^{(n)}_\pi$ needs no modification beyond the well known $s-$ and
$t-$channel contributions.  The values for
the electric and magnetic polarizabilities of the neutron extracted
from the neutron data are of high precision. They are partly in 
agreement with theoretical  predictions.

\section{Acknowledgements}

{\acknowledgement
  One of the authors (M.I.L.) highly appreciates the hospitality of
  the II.\@ Physikalisches Institut der Universit\"at G\"ottingen where
  part of the work was done. He is also very grateful to R. Machleidt
  for a computer code for the CD-Bonn potential.  }

\clearpage


\begin{appendix}

\begin{table}
  \caption{\label{csfree_p}The energy dependence of the free-proton
           differential cross section at 
$\theta^{\rm LAB}_{\gamma}=136.2^{\circ}$.
           The statistical errors are given. The systematic error amounts
           to 4.4\%.}
  \centering
  \begin{tabularx}{.8\columnwidth}{D{.}{.}{3}XD{.}{.}{1}@{~}c@{~}D{.}{.}{1}X}
    \hline\noalign{\smallskip}
    \multicolumn{1}{c}{$E_\gamma \left(\mathrm{MeV}\right)$} &
    \multicolumn{5}{c}{$\frac{d\sigma}{d\Omega_\gamma}
      \left( \frac{\text{nb}}{\text{sr}}\right)$} \\
    \noalign{\smallskip}\hline\noalign{\smallskip}
    211.1 &&  44.4 & $\pm$ &  7.1 & \\
    233.8 &&  59.6 & $\pm$ &  7.6 & \\
    254.3 &&  83.9 & $\pm$ &  8.9 & \\
    273.5 && 122.7 & $\pm$ & 11.1 & \\
    292.7 && 166.1 & $\pm$ & 10.6 & \\
    312.0 && 174.0 & $\pm$ & 11.3 & \\
    331.2 && 150.0 & $\pm$ & 10.0 & \\
    350.4 && 149.1 & $\pm$ & 10.2 & \\
    369.4 && 101.5 & $\pm$ &  8.4 & \\
    388.5 &&  87.6 & $\pm$ &  8.5 & \\
    407.4 &&  83.5 & $\pm$ &  8.5 & \\
    427.3 &&  58.7 & $\pm$ &  7.4 & \\
    448.3 &&  54.1 & $\pm$ &  7.4 & \\
    469.0 &&  55.3 & $\pm$ &  8.1 & \\
    \noalign{\smallskip}\hline
  \end{tabularx}
\end{table}

\begin{table}
  \caption{\label{table_qffr_proton}The energy dependence of the
           differential cross section of "free" proton Compton
           scattering extracted from quasi-free data at
 $\theta^{\rm LAB}_{\gamma}=136.2^{\circ}$. The statistical
           errors are given. The systematic error amounts to 4.4\%.}
  \centering
  \begin{tabularx}{.8\columnwidth}{D{.}{.}{3}XD{.}{.}{1}@{~}c@{~}D{.}{.}{1}X}
    \hline\noalign{\smallskip}
    \multicolumn{1}{c}{$E_\gamma^f \left(\mathrm{MeV}\right)$} &
    \multicolumn{5}{c}{$\frac{d\sigma}{d\Omega_\gamma}
      \left( \frac{\text{nb}}{\text{sr}}\right)$} \\
    \noalign{\smallskip}\hline\noalign{\smallskip}
    231.0 &&  63.7 & $\pm$ & 6.7 & \\
    251.5 &&  77.7 & $\pm$ & 6.5 & \\
    270.6 && 108.7 & $\pm$ & 7.2 & \\
    289.8 && 132.1 & $\pm$ & 6.2 & \\
    309.0 && 149.9 & $\pm$ & 6.6 & \\
    328.2 && 151.2 & $\pm$ & 7.4 & \\
    347.4 && 134.4 & $\pm$ & 8.6 & \\
    366.3 && 106.6 & $\pm$ & 8.6 & \\
    385.4 &&  96.4 & $\pm$ & 8.6 & \\
    \noalign{\smallskip}\hline
  \end{tabularx}
\end{table}

\begin{table}
  \caption{\label{table_qf_proton}The energy dependence of the triple
           differential cross section of the reaction
           $d(\gamma,\gamma^{\,\prime}p)n$ in the center
           of pQFP at $\theta^{\rm LAB}_{\gamma}=136.2^{\circ}$. 
The statistical
           errors  are given. The systematic error amounts to 4.4\%.}
  \centering
  \begin{tabularx}{.8\columnwidth}{D{.}{.}{3}XD{.}{.}{1}@{~}c@{~}D{.}{.}{1}X}
    \hline\noalign{\smallskip}
    \multicolumn{1}{c}{$E_\gamma \left(\mathrm{MeV}\right)$} &
    \multicolumn{5}{c}{$\left( \frac{d^3\sigma}{d\Omega_\gamma
          d\Omega_p dE_p} \right)^\text{CpQFP}
      \left( \frac{\text{nb}}{\text{MeV sr}^2}\right)$} \\
    \noalign{\smallskip}\hline\noalign{\smallskip}
    233.8 &&  49.5 & $\pm$ &  5.1 & \\
    254.3 &&  67.2 & $\pm$ &  5.6 & \\
    273.5 && 103.2 & $\pm$ &  6.8 & \\
    292.8 && 137.4 & $\pm$ &  6.4 & \\
    312.0 && 170.0 & $\pm$ &  7.4 & \\
    331.2 && 185.5 & $\pm$ &  9.0 & \\
    350.4 && 177.2 & $\pm$ & 11.3 & \\
    369.5 && 150.3 & $\pm$ & 12.1 & \\
    388.5 && 144.7 & $\pm$ & 12.8 & \\
    \noalign{\smallskip}\hline
  \end{tabularx}
\end{table}

\begin{table}
  \caption{\label{table_qf_neutron}The energy dependence of the triple
           differential cross section of the reaction
           $d(\gamma,\gamma^{\,\prime}n)p$ in the center of nQFP at
$\theta^{\rm LAB}_{\gamma}=136.2^{\circ}$. The statistical errors are
           given. The systematic error  amounts  to 9.0\%.}
  \centering
  \begin{tabularx}{.8\columnwidth}{D{.}{.}{3}XD{.}{.}{1}@{~}c@{~}D{.}{.}{1}X}
    \hline\noalign{\smallskip}
    \multicolumn{1}{c}{$E_\gamma \left(\mathrm{MeV}\right)$} &
    \multicolumn{5}{c}{$\left( \frac{d^3\sigma}{d\Omega_\gamma
          d\Omega_n dE_n} \right)^\text{CnQFP}
      \left( \frac{\text{nb}}{\text{MeV sr}^2}\right)$} \\
    \noalign{\smallskip}\hline\noalign{\smallskip}
    211.1 &&  17.1 & $\pm$ &  6.3 & \\
    230.2 &&  34.9 & $\pm$ &  8.1 & \\
    249.4 &&  46.9 & $\pm$ &  9.8 & \\
    268.7 &&  67.6 & $\pm$ & 12.8 & \\
    287.9 && 135.7 & $\pm$ & 13.4 & \\
    307.2 && 159.6 & $\pm$ & 15.3 & \\
    326.4 && 217.3 & $\pm$ & 19.8 & \\
    345.6 && 202.2 & $\pm$ & 25.9 & \\
    376.5 && 208.4 & $\pm$ & 23.4 & \\
    \noalign{\smallskip}\hline
  \end{tabularx}
\end{table}

\begin{table}
  \caption{\label{table_qffr_neutron}The energy dependence of the
           differential cross section of free neutron Compton
           scattering extracted from the quasi-free data at
 $\theta^{\rm LAB}_{\gamma}=136.2^{\circ}$. The statistical errors are
           given. The systematic  error  amounts  to 9.0\%.}
  \centering
  \begin{tabularx}{.8\columnwidth}{D{.}{.}{3}XD{.}{.}{1}@{~}c@{~}D{.}{.}{1}X}
    \hline\noalign{\smallskip}
    \multicolumn{1}{c}{$E_\gamma^f \left(\mathrm{MeV}\right)$} &
    \multicolumn{5}{c}{$\frac{d\sigma}{d\Omega_\gamma}
      \left( \frac{\text{nb}}{\text{sr}}\right)$} \\
    \noalign{\smallskip}\hline\noalign{\smallskip}
    208.4 &&  25.7 & $\pm$ &  9.5 & \\
    227.4 &&  46.3 & $\pm$ & 10.8 & \\
    246.6 &&  55.6 & $\pm$ & 11.7 & \\
    265.8 &&  72.5 & $\pm$ & 13.8 & \\
    285.0 && 132.5 & $\pm$ & 13.0 & \\
    304.3 && 142.8 & $\pm$ & 13.7 & \\
    323.4 && 179.4 & $\pm$ & 16.4 & \\
    342.6 && 155.2 & $\pm$ & 19.9 & \\
    373.4 && 143.7 & $\pm$ & 16.2 & \\
    \noalign{\smallskip}\hline
  \end{tabularx}
\end{table}

\end{appendix}

\clearpage
\clearpage


\end{document}